\documentclass[prb,preprint,draft,amsmath,showpacs]{revtex4}
\usepackage{graphicx}

\begin{document}
\DeclareGraphicsExtensions{.eps,.jpg}


\bibliographystyle{prsty}
\input epsf
\title{Far infrared properties of the rare-earth scandate DyScO$_3$}
\author{ L. Baldassarre, A. Perucchi}
\affiliation{ Sincrotrone Trieste S.C.p.A., S.S. 14 km 163.5, in Area Science Park, 34012 Basovizza (Trieste), Italy}\
\author{S. Lupi}
\affiliation{CNR-IOM and Dipartimento di
Fisica, Universit\'{a} di Roma Sapienza, P.le A.Moro 2, 00185
Rome, Italy}\
\author{P. Dore}
\affiliation{CNR-SPIN and Dipartimento di
Fisica, Universit\'{a} di Roma `Sapienza, P.le A.Moro 2, 00185
Rome, Italy}
\begin{abstract}
We present reflectance measurements in the infrared region on a single crystal 
the rare earth scandate DyScO$_3$. Measurements performed between room temperature 
and 10 K allow to determine the frequency of the infrared-active phonons, never 
investigated experimentally, and to get information on their temperature 
dependence. A comparison with the phonon peak frequency  
resulting from \emph{ab initio} computations is also provided. We finally report detailed 
data on the frequency dependence of the complex refractive index of DyScO$_3$ in 
the terahertz region, which is important in the analysis of terahertz 
measurements on thin films deposited on DyScO$_3$. 
\end{abstract}

\maketitle

\section{Introduction}

DyScO$_3$, together with other rare-earth scandates ($R$ScO$_3$), has recently
received considerable attention, since it is considered to be among the best
substrates for the epitaxial growth of high-quality ABO$_3$ perovskite-type thin
films \cite{thin_films}. On such thin films it is possible to induce
ferroelectric and multiferroic properties tailoring their lattice constants by
changing $R$ in $R$ScO$_3$. For example, SrTiO$_3$ exhibits strain-induced
ferroelectricity if grown on a RScO$_3$ substrate\cite{haeni}. Furthermore, scandates
are considered to be some of the most promising candidates to substitute SiO$_2$
as gate dielectric in MOSFET, thanks to 
the high value of their static dielectric constant $\epsilon_0$ 
\cite{haeni, delugas}. SrTiO$_3$/DyScO$_3$ heterostructures are also widely used for
applications in the terahertz (THz) range \cite{kuzel}.

We remark that scandates are increasingly used as substrates for film growth and that the optical investigation of a film often allows basic studies which can be difficult when
only small size single crystals are available. In particular, the large and
flat surface of a film permits accurate optical measurements in the far-IR and THz regions, which have an
important role in studying superconducting films \cite{Tink}.
When the radiation penetration depth is larger than the film thickness
the optical response of the substrate affects 
the measured far-IR/THz spectrum, and the complex dielectric function 
$\tilde{\epsilon}(\omega)=\epsilon_{1}(\omega)+i\epsilon_{2}(\omega)$ 
of the film material 
cannot be obtained through the Kramers-Kronig (KK) transformations. 
In this case more elaborate procedures \cite{Dress, paolod-old} 
must be used to extract the $\tilde{\epsilon}(\omega)$
of the film from the reflectance or transmittance data, which require the knowledge of the substrate complex refractive index 
$\tilde{n}(\omega)=n(\omega)+ik(\omega)$.
While the far-infrared properties of common perovskites-like substrates as 
SrTiO$_3$ and LaAlO$_3$ are well known \cite{calvani, zhang, nkSTO}, 
no far-IR data are available in the 
DyScO$_3$ case. For this system, theoretical calculations
of the phonon modes have been reported \cite{delugas}, 
which could be compared only with data from recent Raman 
investigations\cite{kreisel,kreisel2}. 

Moreover, both infrared (IR) and Raman spectroscopy are of interest in investigating the structural properties of oxide materials with the perovskite structure, 
since the study of the optical phonons can provide direct information on even
subtle structural distortions of the ideal perovskite structure.  

We have performed reflectance measurements in the IR region on a DyScO$_3$ 
single crystal, at a number of temperatures in the 10-300 K range. 
The infrared-active phonons of DyScO$_3$  have been investigated for the first time,
allowing a direct comparison with the results of the $ab$ $initio$ calculations \cite{delugas}. 
Moreover the frequency and temperature dependence of the complex refractive index $\tilde{n}(\omega)$ in the 
far-IR/THz region has been obtained. 

\section{Experimental details and results}
Reflectance measurements were performed at near-normal incidence on the 110 surface 
of a DyScO$_3$ (DSO) single crystal. The sample was glued with silver paint, to ensure thermal contact, on an optically black cone \cite{conetto}  mounted on the end of a Helitran cryostat's coldfinger. Such cone was aligned, with the use of three tilting screws, so to have the sample surface perpendicular to the incident radiation. By employing a home-built reflectivity unit, measurements were performed over a broad energy range 
(50$\div$12000 cm$^{-1}$) with a  Michelson interferometer 
at a spectral resolution of 2 cm$^{-1}$. 

To obtain the absolute value of the reflectivity R($\omega$) we employed the overfilling technique \cite{conetto}.  A metallic  film of Au is deposited  $in$ $situ$ on the sample surface and used as reference. This allows not only to prevent thermal disalignment but also to take into account any effect of diffraction due to the sample size or diffusion if the sample surface is not \emph{mirror-like}.

\begin{figure}[h]   
\begin{center}    
\leavevmode    
\epsfxsize=8.5cm \epsfbox {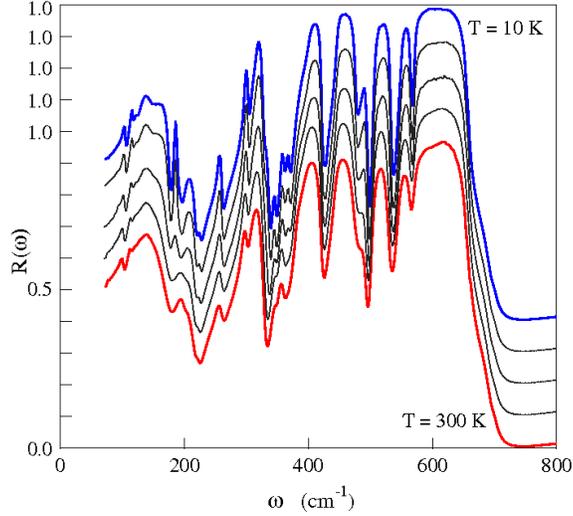}
\caption{\label{rifle}(color online) Reflectivity of DyScO$_3$ shown in the range of far-IR 
for T= 10K, 50K, 150K, 250K and 300K. 
The curves are shifted of 0.1 along the vertical axis for sake of clarity. } 
\end{center}
\end{figure}

\begin{figure}[h]   
\begin{center}    
\leavevmode    
\epsfxsize=8.5cm \epsfbox {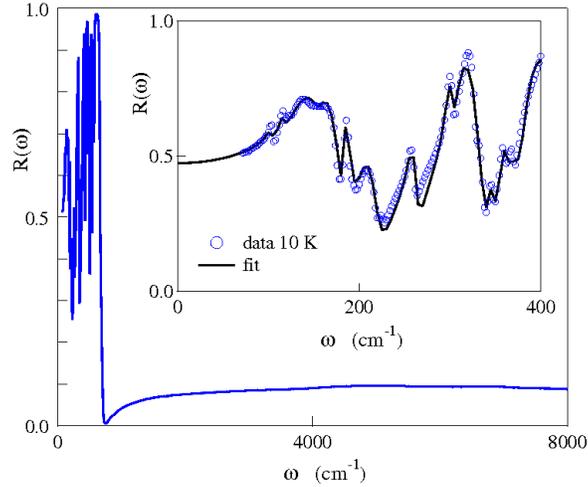}
\caption{\label{rifle2}(color online) Reflectivity of DyScO$_3$ at T= 10 K shown up to 8000 cm$^{-1}$. In the inset the low-frequency R$(\omega)$ is compared with a fit obtained from the Lorentz model.} 
\end{center}
\end{figure}

The optical reflectivity R$(\omega)$ is plotted up to 800 cm$^{-1}$ in Fig. \ref{rifle} at various temperatures.
Several phonon lines, which show a weak temperature dependence, can be observed in this spectral region. At higher frequencies R($\omega$), in agreement with the insulating properties of the DyScO$_3$ crystal, is nearly flat and constant at a value  
of about 0.1 up to the highest measured frequency (see Fig.\ref{rifle2}).

\section{Analysis and discussion}
DyScO$_3$ has an orthorhombically-distorted perovskite structure with space group $Pnma$ with a 20-atom primitive cell. 
Therefore, one expects, as in the isostructural 
LaMnO$_3$ manganite \cite{iliev, fedorov, smirnova, paolone}, 60 phonons, among which 24 are Raman active (corresponding to the irriducible representations $7A_g$, $5B_{1g}$, $7B_{2g}$, $5B_{3g}$), 
25 are infrared active ($9B_{1u}$, $7B_{2u}$, $9B_{3u}$), 8 silent ($8A_u$), and 3 acoustic ($B_{1u}$, $B_{2u}$, $B_{3u}$).
By measuring the reflectance on the 110 surface, we should detect all the $B_{1u}$ modes \cite{kreisel,iliev, fedorov, smirnova} 
corresponding to the dipole moment oscillating along $z$, where the orthorhombic axis $z$ is defined as  $z=[001]$ (see Ref.\cite{kreisel} for more details), and a linear combination of the $B_{2u}$ and $B_{3u}$ corresponding to the dipole oscillating along $y$ and $x$, respectively \cite{fedorov}. 

In order to perform the KK
analysis \cite{Dress}, we first performed a Lorentz fit on 
R$(\omega)$ so to extrapolate its 
behavior as $\omega \rightarrow 0$.  
We have fitted the low-frequency part of  
R$(\omega)$ (\emph{i.e.} below 400 cm$^{-1}$) in order to obtain a careful description in the THz region. This procedure resulted in a convincing extrapolation of the experimental data
down to zero frequency and in reliable results in the far-IR once the KK transformations are performed. We remark that data have been collected up to 12000 cm$^{-1}$, so to avoid problems in the KK procedure due to the high-frequency extrapolation of the spectra.
The resulting $\epsilon_{2}(\omega)$ are reported in Fig. 2 for frequencies below 600 cm$^{-1}$ at selected temperatures. 

At 10 K, 19 of the expected 25 IR-active modes are clearly visible in the spectrum.
As in the case of the isostructural 
LaMno$_3$ system, the three phonon modes (external, bending, and stretching modes) 
proper of the cubic perovskite structure \cite{Burns} are split in several phonons because of the orthorombic distortion. The low frequency phonon modes originate from the external mode, i.e. are due to vibrations of the Dy-cation sublattice with respect to the network of ScO$_6$ octahedra. The high frequency phonons originate from the stretching mode of oxygens in the ScO$_6$ octhaedra, the phonons at intermediate frequencies from the
bending mode. However, due to the strong orthorhombic lattice distortions, none of the modes can be classified as purely bending or as purely stretching as these modes imply considerable changes of both the bond angles and bond lengths in the ScO$_6$ octahedra \cite{fedorov}.
In the vibrational spectra 
we distinguish one phonon mode around 100 cm$^{-1}$ and one around 115 cm$^{-1}$. 
These frequencies are close to those predicted by \emph{ab initio} calculations for a $B_{3u}$ and a B$_{1u}$ modes, respectively. Their dipole moment is due to small displacements of Sc-O ions $vs$ Dy that do not compensate along the $x$ direction\cite{delugas} 
while the larger vertical ($z$) displacements produce no dipole moment. The intensity of these modes can in principle be enhanced by disorder, which can also increase the static dielectric constant $\epsilon_0$.

\begin{figure}[h]   
\begin{center}    
\leavevmode    
\epsfxsize=8.5cm \epsfbox {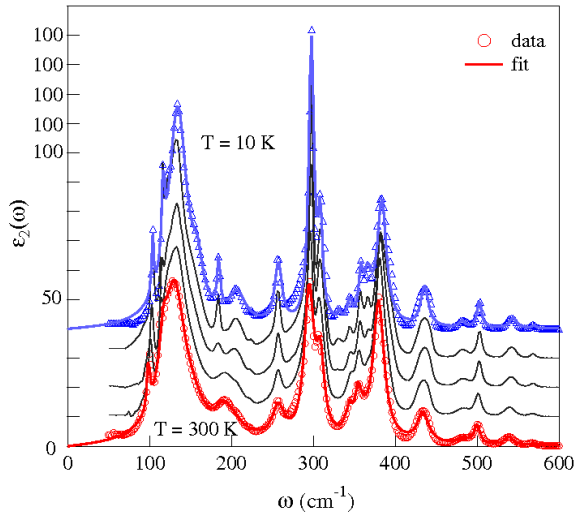}
\caption{\label{fig3}(color online): Imaginary part $\epsilon_2(\omega)$ of the complex
dielectric function of DyScO$_3$. Data are shown at a number of temperatures and shifted for sake of clarity. At 10 K and 300 K both data and Lorentz fitting are reported.}
\end{center}
\end{figure}

To determine the central frequency of the phonon modes and their temperature dependence, we have fitted $\epsilon_2(\omega)$ within the Lorentz model by using:
\begin{equation} \tilde{\epsilon}=\epsilon_1(\omega) + i\epsilon_2(\omega) =  \epsilon_\infty+\sum_j \frac{A_j{\omega_j}^2}{{\omega_j}^2-\omega^2-i\omega\gamma_j} \label{lorentz}\end{equation} where $A_j$, $\omega_j$ and $\gamma_j$ are respectively the mode strength, the central frequency and the damping of each phonon mode $j$. The resulting curves obtained with $\epsilon_\infty\sim 4$  are in good agreement with the data,
as shown for T= 10 K and 300 K in Fig.\ref{fig3}.
The phonon frequencies resulting from the fits are reported in Table \ref{table1} 
and tentatively compared with the calculations of Ref.\cite{delugas}
As our measurements were performed without polarizing the incoming beam
no information about the polarization of the phonon peaks can be extracted 
from the
experimental data. We made a tentative assignment by simply comparing
the phonon frequencies obtained from the fit with the computed ones. Taking
into account the complexity of this class of materials, we believe that the overall
agreement between theory and experiment should be considered as reasonable.

Most of the phonons harden their central frequency with decreasing T and reduce their half-height half-width $\gamma$, as expected and in good agreement with the behavior found in Ref.\cite{kreisel2} at higher temperatures. 
A closer inspection of data shows that 
the broad peak, visible at 300 K just below 200 cm$^{-1}$ is due to two different phonon modes at 190 cm$^{-1}$ and at 199 cm$^{-1}$ (see Table \ref{table1}). 
The apparent splitting of such peak is due to the 
softening of the more intense mode (at 190 cm$^{-1}$) to 185 cm$^{-1}$ 
and the simultaneous hardening of the second mode. 
Some other weak modes also show a light softening by 
decreasing T as reported in Table.\ref{table1}.
This softening might correspond to subtle lattice modifications (still within the $Pnma$ crystal structure) as T decreases. To the best of our knowledge, the T-behavior of the DyScO$_3$ lattice parameters has been investigated between 298 - 1273 K,  that is the temperature range of 
interest for perovskite thin film growth\cite{lattice}, while no data exist at lower temperatures. Therefore one cannot link unambiguously the modifications in the IR-active phonon response to a structural rearrangement.

\begin{table}
\caption{Experimental phonon frequencies obtained by means of IR measurements (at 300 K 
and 10 K) compared with theoretical calculations, from Ref.\cite{delugas}} \label{table1}
\begin{center}
\vspace{0.5cm}
\begin{tabular}{ p{.08\textwidth}p{.08\textwidth}p{.08\textwidth}p{0.08\textwidth} p{0.095\textwidth} }
\hline \hline
 $\omega_{exp}$ & $\omega_{exp}$ &$A_j$ & $\omega_{theo}$  &  Mode\\
 (cm$^{-1}$) & (cm$^{-1}$)&  & (cm$^{-1})$ & Symmetry \\
 \hline
300 K & 10 K & 10 K & Ref.\onlinecite{delugas}&Ref.\onlinecite{delugas}\\
\hline
101 & 104 & 0.52& 97.9 & B$_3u$\\
113& 117 & 0.96&111.0&B$_1u$\\
131&135& 15.35 &129.4&B$_2u$\\
160& 159 & 1.12& 173.7&B$_1u$\\
190& 185 & 0.37& 193.3&B$_2u$\\
199& 205&1.14 &197.0&B$_3u$\\
- & - &-& 231.4&B$_3u$\\
256 & 257 & 0.85 & 278.8&B$_1u$\\
- & - & - &283.2&B$_2u$\\
- & - & -& 285.3&B$_3u$\\
295&298&2.53&293.9&B$_1u$\\
306& 309 & 0.69&328.8&B$_1u$\\
344&332& 0.04& 334.4&B$_3u$\\
346&344& 0.14& 336.4&B$_2u$\\
354&358&0.23&356.7&B$_3u$\\
371&367&0.44&368.6&B$_2u$\\
381&385&1.51&369.5&B$_1u$\\
- & - & -& 412.9&B$_1u$\\
- & - & -&419.2&B$_3u$\\
434&436&0.35&435.5&B$_2u$\\
- & - & -&445.3&B$_2u$\\
- & - & -&478.1&B$_3u$\\
484&482&0.03&484.8&B$_1u$\\
501&503&0.12&509.7&B$_1u$\\
539&544&0.07&532.6&B$_3u$\\
570&570&0.01&-&\\
  \\
 \hline
\hline
\end{tabular}

\end{center}
\end{table}

It is important to notice that the low-frequency modes, due to vibrations of
the $R$-cation sublattice with respect to the ScO$_6$ octahedral network,
are those with higher intensity. 
This finding supports the theoretical 
prediction\cite{delugas} that in DyScO$_3$ the high value of the static dielectric constant 
$\epsilon_0$, i.e. $\tilde{\epsilon}(\omega\rightarrow0)$, is associated with low-frequency 
ionic vibrations. On the basis of the employed procedure, $\epsilon_0$ is given by 
$\epsilon_\infty+\sum_jA_j$ (see Eq.\ref{lorentz}). 
Since the fitting 
procedure provides the $A_j$ value for each mode (as reported in Table \ref{table1}), 
we find that the phononic contribution to $\epsilon_0$, i.e. $\sum_jA_j$ is large, of about 21, 
mainly due to the phonon mode at 135 cm$^{-1}$.
Moreover we find a total static dielectric constant $\epsilon_0\approx 25$, as the sum
of $\epsilon_\infty\sim4$ and the phononic contribution, in excellent agreement with
theory\cite{delugas} and experiments on single crystals\cite{haeni}.

\begin{figure}[h]   
\begin{center}    
\leavevmode    
\epsfxsize=8.5cm \epsfbox {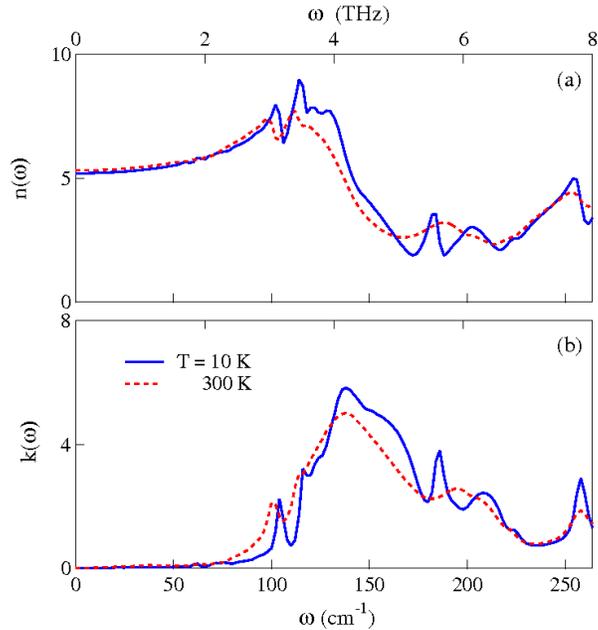}
\caption{\label{fig4}(color online): a) Real and b) Imaginary part of $\tilde{n}$ for DyScO$_3$ at 10 and 300 K in the THz range.}.
\end{center}
\end{figure}

Finally we report in Fig.\ref{fig4} the frequency dependance of the complex refractive
index $\tilde{n}=n+ik$ in the THz region as directly obtained from 
$\tilde{\epsilon}=\epsilon_{1}+i\epsilon_{2}$. 
Below 100 cm$^{-1}$ $n$ is nearly flat reaching a value of about 5 for $\omega \rightarrow 0$, while the vanishingly small $k$ value indicates the absence of an appreciable absorption in the same spectral region. 

The $n$ and $k$ data at 10 K have been recently employed in the analysis of the THz measurements performed on a film\cite{perucchi}, grown on a DSO substrate\cite{JJ}, of the BaFe$_{1.84}$Co$_{0.16}$As$_2$ compound, that belongs to the class of the new Fe-based
superconductors which attracted strong attention since
their recent discovery \cite{pnict1}. 

\section{Conclusions}
We have presented here the first experimental data on the IR phonon spectrum 
of the rare earth scandate DyScO$_3$. In the temperature range between 10 and 300 K no dramatic changes occur in the phonon response indicating that the $Pnma$ structure is stable down to the lowest measured temperature. The overall agreement with the recent \emph{ab initio} calculations is to be considered satisfactory.
We finally note that the present work can provide reference data and extrinsic peak selection for future infrared investigations of thin films grown on DyScO$_3$ substrates.



\begin{thebibliography}{99}
\bibitem{thin_films}  Uecker, R., Velickov,  B.,  Klimm, D.,  Bertram, R.,  Bernhagen, M., Rabe, M., Albrecht, M., Fornari R., and  Schlom, D.G., \emph{Journal of Crystal Growth}, \textbf{310}, 2649 (2008).

\bibitem{haeni} Haeni, J.H., Irvin, P., Chang,  W.,  Uecker, R., Reiche, P.,  Li, Y.L.,  Choudhury, S.,   Tian, W.,  Hawley, M E., Craigo,B., Tagantsev, A.K., Pan, X.Q.,  Streiffer, S.K., Chen, L.Q.,  Kirchoefer, S.W., Levy, J., and Schlom, D.G., \emph{Nature}, \textbf{430}, 758, (2004).

\bibitem{delugas}  Delugas, P., Fiorentini, V., Filippetti, A.,  and Pourtois, G., \emph{Phys. Rev. B}, \textbf{75},  115126 (2007).

\bibitem{kuzel} Kuzel, P., Kadlec, F.  Petzelt, J., Shubert, J.,  Panaitov, G., \emph{Appl. Phys. Lett.}, \textbf{91}, 232911 (2007). 

\bibitem{Tink}  Tinkham, M., \emph{Introduction to Superconductivity}, 
McGraw-Hill, New York (1975)

\bibitem{Dress}  Dressel, M.,  and Gr\"{u}ner, G.,  \emph{Electrodynamics of Solids},
Cambridge University Press, Cambridge, England (2002). 

\bibitem{paolod-old}  Berberich, P., Chiusuri, M., Cunsolo, S., Dore, P., Kinder, H., Varsamis, C.P., \emph{Infrared Phys.} \textbf{34}, 269 (1993).

\bibitem{calvani}  Calvani, P.,  Capizzi, M.,  Donato, F.,  Dore, P.,  Lupi, S.,  Maselli, P.,  Varsamis, C.P. \emph{Physica C} 181, 289 (1991).

\bibitem{zhang}  Zhang, Z.M., Choi, B.I., Flik, M.I.,  and Anderson, A.C., \emph{J. Opt. Soc. Am. B} {\bf11}, 2252 (1994)

\bibitem{nkSTO} Dore, P., Paolone, A., and Trippetti, R., \emph{J. Appl. Phys.} 80 
5270 (1996);  Dore, P.,  De Marzi, G., and Paolone, A., \emph{Int. J. of IR and mm Waves}, {\bf18}, 125 (1997). 

\bibitem{kreisel}  Chaix-Pluchery, O.,and Kreisel, J., \emph{J. Phys.: Condens. Matter} \textbf{21}, 175901 (2009). 

\bibitem{kreisel2} Chaix-Pluchery, O.,and Kreisel, J., \emph{J. Phys.: Condens. Matter} \textbf{22} 165901 (2010).

\bibitem{conetto}  Homes, C.C.,  Reedik, M.,  Cradles, D.A.,  and Timusk, T.,  \emph{Appl.Optics} \textbf{32}, 2976 (1993).

\bibitem{iliev} Iliev, M.N., Abrashev, M.V., Lee, H.-G., Popov, V.N.,  Sun, Y.Y.
 Thomsen, C., Meng,  R.L. and Chu,  C.W., \emph{Phys. Rev. B} {\bf57}, 2872 (1998).

\bibitem{fedorov}  Fedorov, I.,  Lorenzana, J.,  Dore, P.,  De Marzi, G., Maselli,  P.,  Calvani, P., Cheong, S-W., Koval, S., and  Migoni, R., \emph{Phys. Rev. B} {\bf 60}, 11875 (1999).

\bibitem{smirnova}  Smirnova, I.S., \emph{Physica B} {\bf262}, 247 (1999).

\bibitem{paolone}  Paolone, A., Roy, P., Pimenov, A., Loidl, A., MelÕnikov, O.K.,
and Shapiro,  A.Ya., \emph{Phys. Rev. B }  {\bf61}, 11255 (2000).

\bibitem{Burns}  Burns, G.,  \emph{Solid State Physics} Academic Press, Boston, 1990.

\bibitem{lattice}  Biegalski, M.D., Haeni, J.H., Trolier-McKinstry, S., Schlom,  D.G., Brandle,  C.D.,  and Ven Graitis, A.J.,  \emph{J. Mater. Res.}, {\bf 20}, 952 (2005). 

\bibitem{perucchi} Perucchi, A., Baldassarre, L., Marini, C., Lupi, S., Jiang,  J., Weiss, J.D., Hellstrom, E.E., Lee, S., Bark, C.W.,  Eom, C.B., Putti, M., Pallecchi, I.,  and Dore, P.,  \emph{EPJ B, in press}  (2010).

\bibitem{JJ} Lee, S., Jiang, J., Zhang, Y., Bark, C.W.,  Weiss, J.D., Tarantini, C., Nelson,  C.T.,  Jang, H.W., Folkman, C.M., Baek, S.H.,  Polyanskii, A., Abraimov, D., Yamamoto, A., Park, J.W., Pan, X.Q., Hellstrom, E.E., Larbalestier, D.C., and  Eom, C.B., \emph{Nat. Mat.}, \textbf{9}, 397, (2010).

\bibitem{pnict1} Kamihara, Y.,  Watanabe, T.,  Hirano,  M., and Hosono, H., \emph{J. Am. Chem. Soc.} \textbf{130}, 3296 (2008).


\end{thebibliography}
\end{document}